\documentclass[11pt]{article}

\usepackage{epsfig}

\usepackage{a4}
\usepackage{amssymb}
\long\def\symbolfootnote[#1]#2{\begingroup%
\def\thefootnote{\fnsymbol{footnote}}\footnote[#1]{#2}\endgroup}

\begin{document}

	\newcommand{\BA}          {1}
	\newcommand{\BERN}        {2}
	\newcommand{\BO}     	  {3}
	\newcommand{\JINR}        {4}
	\newcommand{\LNGS} 	  {5} 
	\newcommand{\LYON}        {6}
	\newcommand{\NA}	        {7}
	\newcommand{\NEU}	        {8}
	\newcommand{\ROME}	  {9}
	\newcommand{\SA}	       {10}

	\newcommand{\Institutes}{
	\BA . Dipartimento di Fisica dell'Universit$\rm\grave{a}$ di Bari and INFN, 70126 Bari, Italy \\
	\BERN . University of Bern, CH-3012 Bern, Switzerland \\
	\BO . Dipartimento di Fisica dell'Universit$\rm\grave{a}$ di Bologna and INFN, 40127 Bologna, Italy \\
	\JINR . JINR - Joint Institute for Nuclear Research, 141980 Dubna, Russia \\
	\LNGS . Laboratori Nazionali del Gran Sasso, 67010 Assergi, L'Aquila, Italy \\
	\LYON . IPNL, IN2P3-CNRS and Universit$\acute{e}$ Claude Bernard Lyon, 69622 Villeurbanne, France \\
	\NA . Dipartimento di Fisica dell'Universit$\rm\grave{a}$ ``Federico II'' and INFN, 80125 Napoli, Italy \\
	\NEU . University of Neuch$\hat{a}$tel, CH-2000 Neuch$\hat{a}$tel, Switzerland \\
	\ROME . Dipartimento di Fisica dell'Universit$\rm\grave{a}$ ``La Sapienza'' and INFN, 00185 Roma, Italy \\
 	\SA . Dipartimento di Fisica dell'Universit$\rm\grave{a}$ di Salerno and INFN, 84084 Fisciano, Salerno, Italy \\
	}

	\newcommand{\AuthorList}{L.~Arrabito$^{\LYON}$, C.~Bozza$^{\SA}$, S.~Buontempo$^{\NA}$, L.~Consiglio$^{\BO}$, M.~Cozzi$^{\BO}$, N.~D'Ambrosio$^{\LNGS}$, G.~De~Lellis$^{\NA}$, 
	M.~De~Serio$^{\BA}$\symbolfootnote[2]{Corresponding author. E-mail: Marilisa.Deserio@ba.infn.it}, 
	F.~Di~Capua$^{\NA}$, D.~Di~Ferdinando$^{\BO}$, N.~Di~Marco$^{\LNGS}$, A.~Ereditato$^{\BERN}$, L.~S.~Esposito$^{\LNGS}$, R.~A.~Fini$^{\BA}$, G.~Giacomelli$^{\BO}$, 
	M.~Giorgini$^{\BO}$, G.~Grella$^{\SA}$, M.~Ieva$^{\BA}$, J.~Janicsko~Csathy$^{\NEU}$, F.~Juget$^{\NEU}$, I.~Kreslo$^{\BERN}$, I.~Laktineh$^{\LYON}$, K.~Manai$^{\LYON}$, 
	G.~Mandrioli$^{\BO}$, A.~Marotta$^{\NA}$, P.~Migliozzi$^{\NA}$, P.~Monacelli$^{\LNGS}$, U.~Moser$^{\BERN}$, M.~T.~Muciaccia$^{\BA}$, A.~Pastore$^{\BA}$, 
	L.~Patrizii$^{\BO}$, Y.~Petukhov$^{\JINR}$, C.~Pistillo$^{\BERN}$, M.~Pozzato$^{\BO}$, G.~Romano$^{10}$, G.~Rosa$^{\ROME}$, A.~Russo$^{\NA}$, N.~Savvinov$^{\BERN}$, 
	A.~Schembri$^{\ROME}$, L.~Scotto~Lavina$^{\NA}$, S.~Simone$^{\BA}$, M.~Sioli$^{\BO}$, C.~Sirignano$^{\SA}$, G.~Sirri$^{\BO}$, P.~Strolin$^{\NA}$, V.~Tioukov$^{\NA}$ 
	and T.~Waelchli$^{\BERN}$.}

	\title{Track reconstruction in the emulsion-lead target of the OPERA experiment using the ESS microscope}

	\maketitle

	\author{\noindent \\ \AuthorList }

	\begin{flushleft}
	\footnotesize{\Institutes}
	\end{flushleft}

    \begin{abstract}

    The OPERA experiment, designed to conclusively prove the existence of
    $\rm \nu_\mu \rightarrow \nu_\tau$ oscillations in the atmospheric sector,
    makes use of a massive lead-nuclear emulsion target to observe the appearance
    of $\rm \nu_\tau$'s in the CNGS $\rm \nu_\mu$ beam.
    The location and analysis of the neutrino interactions in \emph{quasi} real-time
    required the development of fast computer-controlled microscopes able to reconstruct
    particle tracks with sub-micron precision and high efficiency at a speed of $\sim 20 \, \rm cm^2 / h$.
    This paper describes the performance in particle track reconstruction of the
    \emph{European Scanning System}, a novel automatic microscope for the measurement
    of emulsion films developed for OPERA.

    \end{abstract}

\section{Introduction}

In the last decades, several experiments using solar, atmospheric, reactor and accelerator neutrinos
have provided unambiguous evidence for neutrino mixing.
Strong indications in favour of the hypothesis of $\nu_\mu \rightarrow \nu_\tau$ oscillations
in the atmospheric sector have been obtained by \cite{SK,Kamio,MACRO,Soudan2} and have been more recently
confirmed by \cite{K2K,MINOS}.
However, the direct observation of the \emph{appearance} of oscillated $\nu_\tau$'s in the atmospheric
$\nu$ signal region of the parameter space is still an open issue.
The OPERA experiment \cite{opera} has been designed to achieve this goal using the
\emph{CERN to Gran Sasso} $\nu_\mu$ beam (CNGS \cite{CNGS}) over a baseline of $730 \, \rm km $.

Based on the Emulsion Cloud Chamber technique \cite{ECC},
nuclear emulsions \cite{emu} are used as sub-micrometric space resolution trackers to detect the
$\tau$ leptons produced in $\nu_\tau$ charged current interactions in a massive ($\rm \mathcal{O} (kt)$)
lead target.

The basic target unit (\emph{brick}) with size $12.7 \times 10.2 \times 7.5 \, \rm cm^3$
consists of $57$ nuclear emulsion films, made of two $44 \, \rm \mu m$-thick layers on either side of
a $205 \, \rm \mu m$ plastic base, interleaved with $56$ lead plates of $1 \, \rm mm$ thickness. 
Bricks are arranged in vertical structures called \emph{walls}.

Electronic trackers, made of planes of plastic scintillator strips inserted in between the target walls, 
and muon spectrometers accomplish the task of identifying in real-time the position of the neutrino interaction 
inside the target. As a result of the combined analysis of electronic data, a 3-D brick probability map is 
computed for each single event and the brick with the highest probability is promptly removed from the target 
during the run. The so-called \emph{Changeable Sheets} (CS),
pairs of emulsion films placed in front of each brick and acting as interfaces with the electronic trackers,
allow the brick identification signal to be checked. Only in case of confirmation, the emulsion films of the brick 
are developed and sent to the scanning laboratories for interaction vertex location and further analysis.

Assuming the nominal intensity of the CNGS beam ($4.5 \times 10^{19}$ p.o.t. / year), about $30$ bricks
per day will be extracted from the target, corresponding to a total surface of nuclear emulsions
to be scanned of a few thousands $\rm cm^2$ per day\footnote{The position resolution of the 
electronic trackers ($\sim 1 \, \rm cm$) will allow to reduce the area to be scanned to $\sim 25 \, \rm cm^2$ 
for CC interactions with penetrating muon tracks. For NC events, we foresee to scan 
the whole surface.}
In order to analyze the events in \emph{quasi} real-time
with a reasonable number of microscopes \mbox{($\sim \,$ 1 microscope / brick / day)}, the use of fast automatic systems 
with a scanning speed of about $20 \, \rm cm^2 / h$ is required.

Moreover, the detection of  the short-lived $\tau$ leptons through the observation of their \emph{kink}
decay topology, namely an abrupt deviation of the $\tau$ particle trajectory at the decay point in
electron, muon and charged hadron(s), occurring over distances of $\sim 1 \, \rm mm$ at CNGS energy, 
demands adequate position ($\stackrel{<}{\sim} 1 \, \rm \mu m$) and angular ($\sim 1 \, \rm mrad$) resolutions.

The \emph{European Scanning System} (ESS) was designed to meet these experimental requirements\footnote{A
different system for the scanning of OPERA films at high speed has been developed in Japan (SUTS \cite{SUTS}).}.
The general structure of the system and the algorithms developed for image processing and particle tracking
in single emulsion films, as well as the hardware and its performance, have been already described
in detail \cite{ESS,ESS_HW,ESS_PRECISION}. 
The performance in electron-pion separation has been recently published in \cite{ELEPI}.

In this paper, after a brief summary of the main features of the ESS, the film-by-film alignment algorithm,
relevant for the analysis of OPERA bricks, and the tracking performance of the system are presented.
The methods to be applied for the location of neutrino events are then described and
the results of a beam exposure to high-energy pions, designed to test the above procedures, are reported.

\section{The European Scanning System}

\begin{figure}
\begin{center}
\epsfig{file=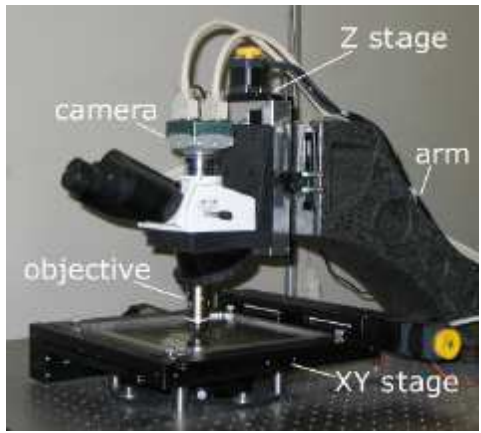}

\caption{The \emph{European Scanning System} microscope. \label{fig:ESS}}
\end{center}
\end{figure}
The ESS microscope, shown in Fig. \ref{fig:ESS}, consists of

\begin{itemize}

\item computer driven horizontal and vertical stages equipped with high-speed precision mechanics;

\item customized optics providing achromatic planar images of the emulsion;

\item high resolution and high frame-rate camera interfaced with a programmable frame-grabber and vision processor.

\end{itemize}

By adjusting the focal plane of the objective lens through the whole emulsion thickness,
a sequence of tomographic images of each field of view are taken at equally spaced depth levels, 
processed and analysed in order to recognise aligned clusters of dark pixels (\emph{grains}) produced
by charged particles along their trajectories. 
Each $44 \, \rm \mu m$-thick OPERA emulsion layer is spanned by 15 tomographic images in steps of $\sim 3 \, \rm \mu m$, 
accounting for the effective focal depth of the system.

\begin{figure}[t]
\begin{center}
\epsfig{file=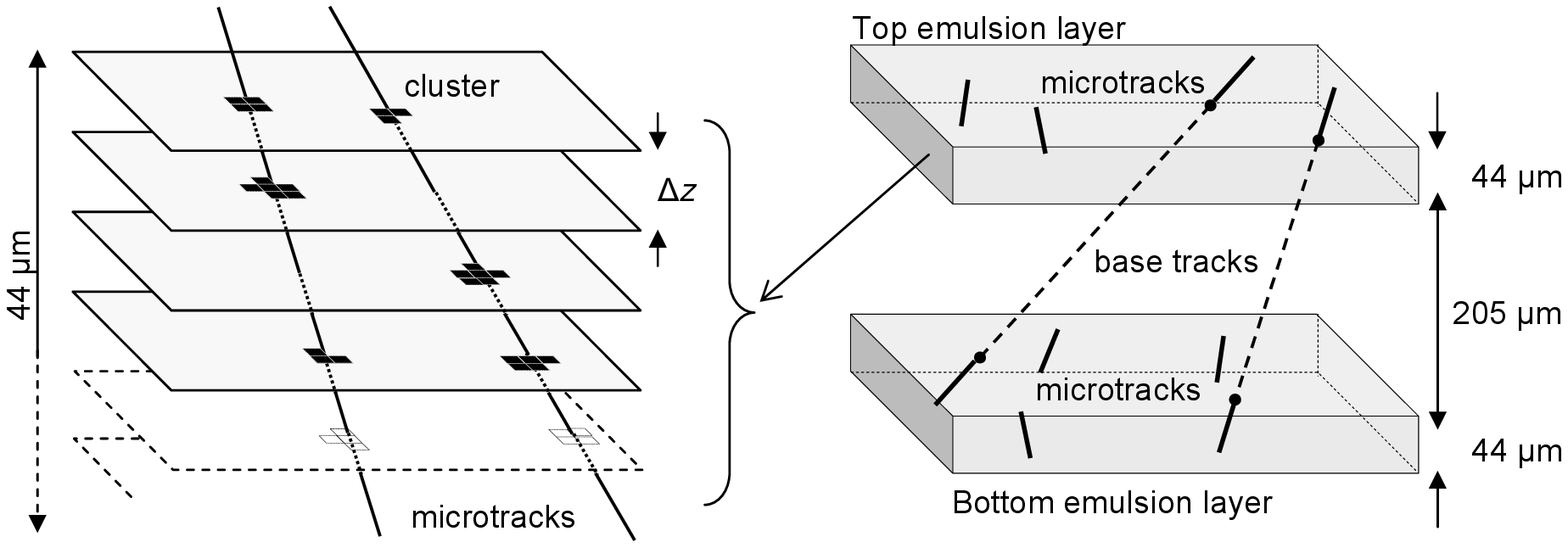,scale=0.75}

\caption{Micro-track connection across the plastic base (base-track). \label{fig:basetrack_scheme}}
\end{center}
\end{figure}

\emph{Micro-tracks} are reconstructed as geometrical alignments of grains detected in
different levels within the same layer. By connecting micro-tracks across the plastic support, 
the so-called \emph{base-tracks} are formed (Fig. \ref{fig:basetrack_scheme}). This strongly reduces
the instrumental background due to fake combinatorial alignments, thus significantly improving 
the signal to noise ratio, and increases the precision of track angle reconstruction by
minimising distortion effects.

The reconstruction of particle tracks in OPERA bricks requires connecting base-tracks in several
consecutive films.

\section{Emulsion film alignment}

In order to define a global reference system (hereafter referred to as \emph{brick} reference system),
prior to track reconstruction a set affine transformations (shift, rotation and expansion)
relating track coordinates in consecutive films have to be computed to account for relative misalignments
and deformations.
The mechanical accuracy of film piling in brick assembly is indeed of $50 \div 100 \, \rm \mu m$, much worse
than the achievable precision. Furthermore, emulsion films are affected by environmental conditions
(temperature and humidity) altering their original geometry.
The task of film alignment is accomplished by exposing each brick to a controlled flux of cosmic rays
before disassembly, as described in \cite{cosmics}, and by applying the following procedure film by film.

Starting from a set of independent measurements in single emulsion films, an aligned volume is created
through an iterative pattern matching procedure computing the parameters of the transformations

\begin{center}
$    \left( \begin{array}{c} x^{brick} \\ y^{brick} \end{array} \right) \, =
     \left( \begin{array}{cc} a_{11} & a_{12} \\ a_{21} & a_{22} \end{array} \right) \,
     \left( \begin{array}{c} x^{film} \\ y^{film} \end{array} \right) \, +
     \left( \begin{array}{c} b_1 \\ b_2 \end{array} \right) \, ,$
\end{center}
where $x^{film}$, $y^{film}$ are single film track coordinates
and $x^{brick}$, $y^{brick}$ are the corresponding \emph{aligned} ones.
A least square fit is applied after maximising the number of matching pairs within predefined position and
slope tolerances measured in three zones, typically chosen at the corners of the scanned area
in order to maximise the lever arm and thus disentangle the contributions due to rotation and translation.

Moreover, a wide angular spectrum of collected tracks allows the total thickness of the emulsions and 
of the interleaved lead plates to be precisely computed, accounting for possible a-planarities. 
The difference between the real position of a given track in the \emph{i-th} film and the extrapolated position
from the \emph{j-th} film depends indeed on the uncertainty $\rm \delta Z$ on the knowledge of
the thickness and is a linear function of the track angle.
By using a set of tracks crossing the brick at different inclination angles, a fit can be performed and
the correction to be applied to the nominal value of the Z coordinate, perpendicular to the emulsion plane,
can be estimated, thus improving the quality of the alignment.

By applying the above procedure to each pair of consecutive films, relative displacements can be reduced
to the level of a few $\rm \mu m$, adequate for film-by-film track following (see Section 5), down to less than
$1 \, \rm \mu m$, as required by local-scale ($\sim 1 \, \rm mm$) analysis (this subject will be discussed in forthcoming papers). 

The achievable precision mainly depends on the number of penetrating tracks, proportional to the area of
the measured zones.
The density of passing-through tracks should be low enough in order not to spoil the topological and kinematical reconstruction
of neutrino events; on the other hand, the scanning time is a critical issue and needs to be minimised.
Typically, a density of the order of a few tracks $/ \rm mm^2$ and scanning zones of
$\rm \mathcal{O} (10 \, mm^2)$ are a reasonable compromise between these two conflicting requirements.

Once all films have been aligned, base-tracks are connected to form \emph{volume tracks}.
A two-step algorithm is applied \cite{Fedra}. The first step consists of identifying incremental \emph{chains} of consecutive base-tracks.  
These chains are then used as input for the Kalman filter algorithm \cite{Kalman}, performing track fitting and propagation 
with a maximum number of allowed consecutive \emph{holes} (i.e. missing segments) accounting for tracking inefficiency.

\section{Study of the ESS basic performance}

In order to study
the track reconstruction performance of the ESS in terms of resolution and efficiency,
a test exposure was performed at the CERN PS in the T7 experimental area. One brick consisting of
$64$ emulsion films with no lead plates in between was exposed
to $10 \, \rm GeV/\it c$ negatively-charged pions. The choice of high beam momentum and the use of emulsions only 
were motivated by the need to minimise the effect of multiple Coulomb scattering that would spoil the measurement 
of the intrinsic resolution.

The emulsion films, produced by the Fuji Film\footnote{Fuji Film, Minamiashigara, 250-0193, Japan.} company,
were similar to those used for the experiment, although they were not part of the official OPERA production.
They were transported by plane from Japan to CERN where the \emph{refreshing} procedure was applied shortly before exposure. \\
Nuclear emulsions are continuously sensitive to charged particles from production time until development.
As a consequence, they normally accumulate a significant amount of background due to cosmic rays and
ambient radioactivity that cannot be distinguished from particle tracks originating from the beam.
For this reason, a process of controlled fading of the emulsions, called \emph{refreshing}, consisting in
the destruction of latent image centers in silver bromide crystals by keeping the films at high temperature
and high relative humidity for a few days, was implemented for the OPERA experiment \cite{REFRESH}.
The working conditions can be tuned in order to reduce the average number of grains per base-track below
the applied threshold, while preserving the emulsion sensitivity. This implies the \emph{cancellation} of
recorded tracks. The achieved track erasing efficiency can be higher than $98 \%$.

The angular dependence of the ESS tracking efficiency and precision were studied by rotating the brick
with respect to the beam direction in 7 different positions from $0 \, \rm mrad $ (tracks perpendicular to
the films) to $600 \, \rm mrad$.
The pion beam and the exposure time were adjusted so as to produce a density in emulsion of about
$3$ tracks $/ \rm mm^2$ for each angle.
Reference emulsions not exposed to the pion beam were also developed to provide an evaluation of the
instrumental background, mainly due to uncorrelated grains randomly produced in the development process
(\emph{fog}) that tend to hide real track grains, hence resulting in fake coincidences.

In order to collect a sample of several hundred tracks per angle exposure, an area of a few $\rm cm^2$ was
scanned for each film and a pattern-matching procedure was applied to compute the parameters of
film inter-calibration transformations, as described in Section 3. Measured base-tracks were then connected
to form volume tracks.

\begin{figure*}[t!]
\begin{center}
\begin{tabular}{cc}
    \epsfig{file=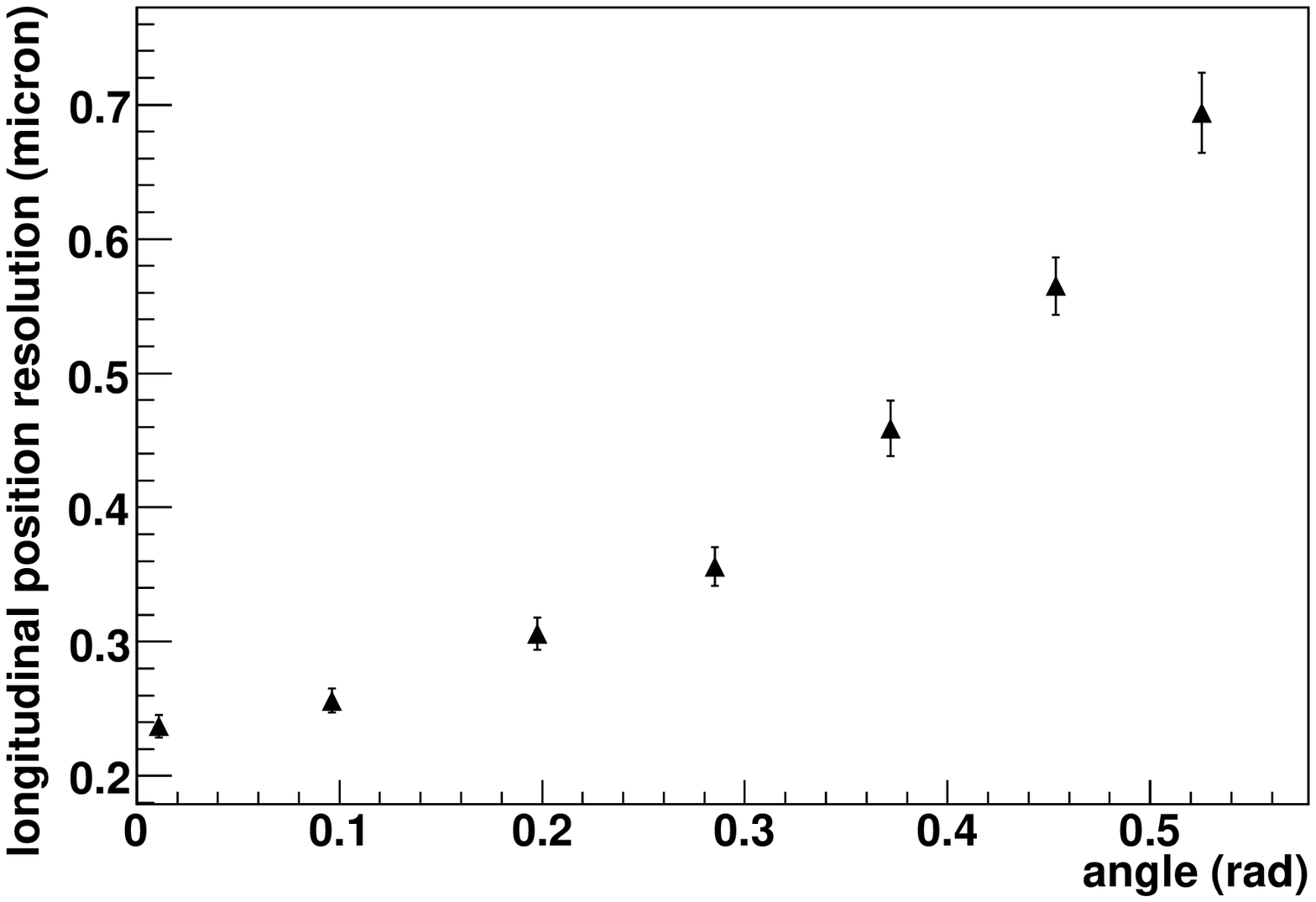,scale=0.35}
        & \epsfig{file=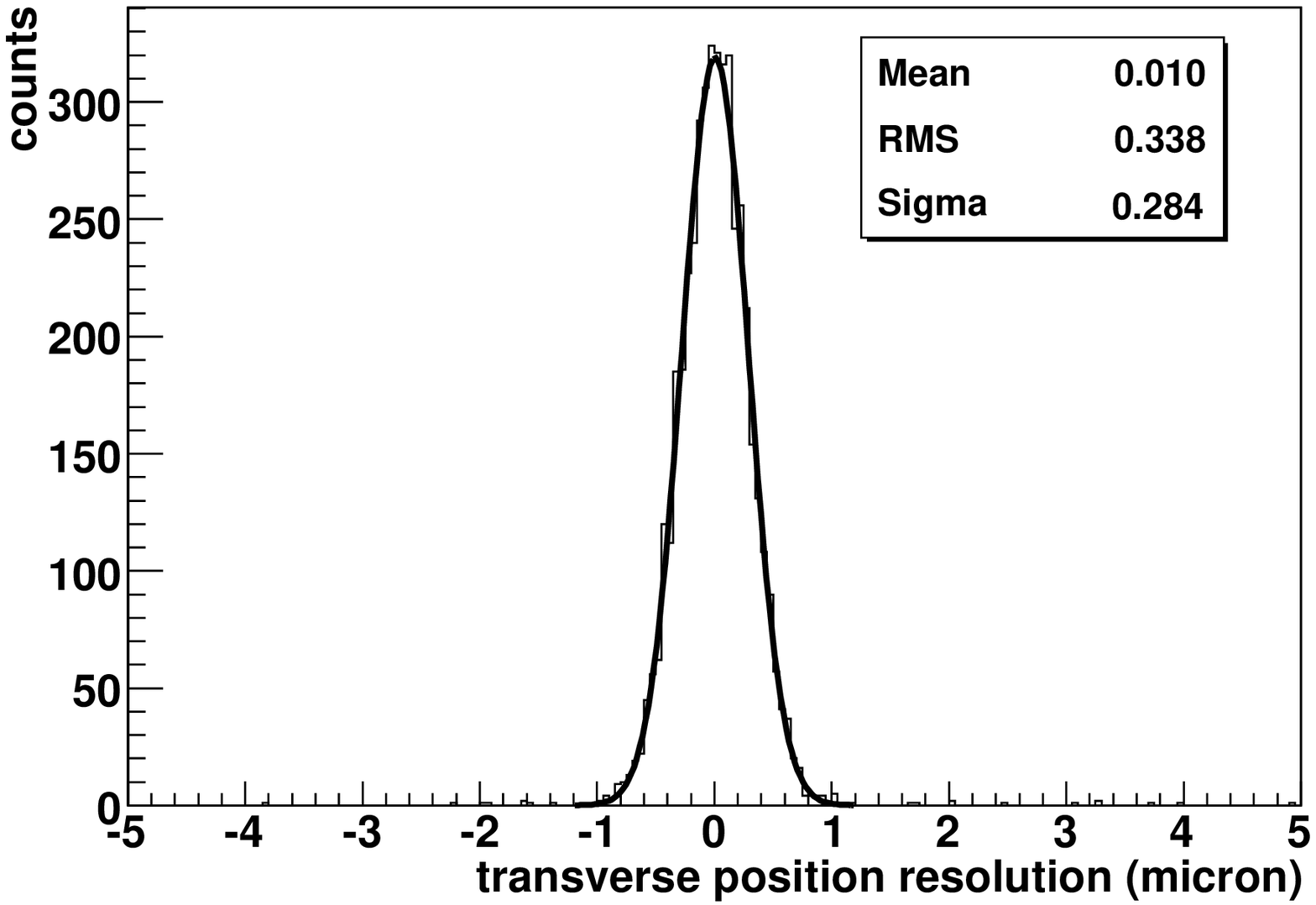,scale=0.35} \\
\end{tabular}

\caption{Longitudinal position residuals between base-tracks and corresponding volume tracks as a function of track angle (left);
transverse position residuals (right). \label{fig:vt_pos}}
\end{center}
\end{figure*}

Fig. \ref{fig:vt_pos} shows the position residuals between each base-track and the corresponding
fitted volume track as a function of the track angle. If $\bf{S} = (s_x, s_y)$ is the slope vector of a straight line
representing the projection of a particle track in the horizontal plane, the unit vectors
\mbox{$\bf{n_\parallel}$ $\, = \, (n_x,n_y)$} such that $\bf{S}$ $\, = \, S \, \bf{n_\parallel}$ and its normal
\mbox{$\bf{n_\perp}$ $\, = \, (n_y,-n_x)$} define two orthogonal directions, called \emph{longitudinal} and
\emph{transverse}, that can be used to decouple the slope-dependent contribution from the intrinsic
accuracy.
The longitudinal component is indeed affected by a slope-dependent term proportional to the uncertainty 
on the Z positions of grains that can be assumed to be given by the sampling step of the tomographic sequence \cite{ESS}. 

The left plot of Fig. \ref{fig:vt_pos} shows longitudinal residuals ranging from $0.3 \, \rm \mu m$ to
$0.7 \, \rm \mu m$.
Transverse residuals (right plot) are independent of track angle and represent the ultimate precision
($0.3 \, \rm \mu m$) that can be achieved accounting for measurement errors.

\begin{figure*}
\begin{center}
\begin{tabular}{cc}
   \epsfig{file=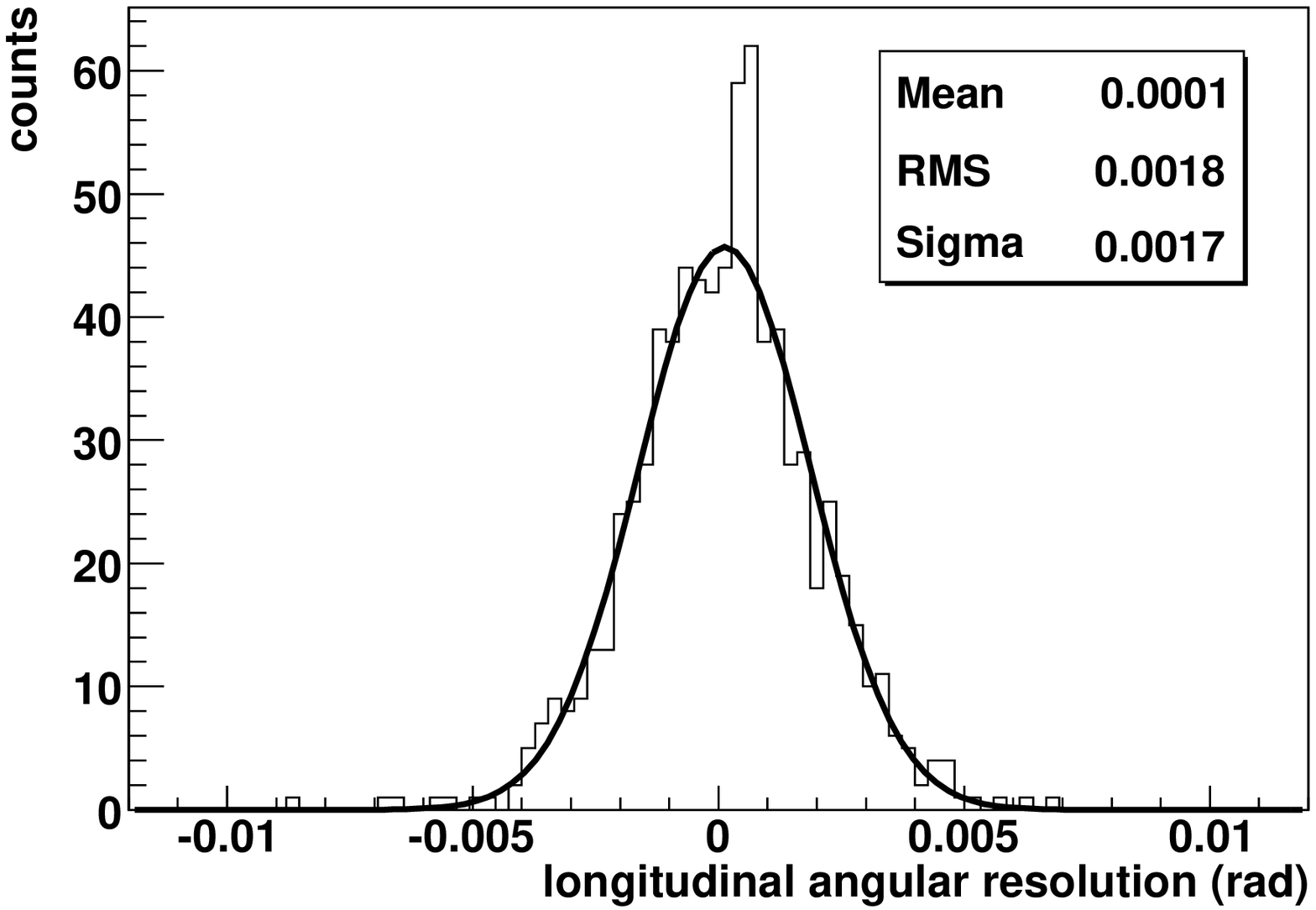,scale=0.35}
        & \epsfig{file=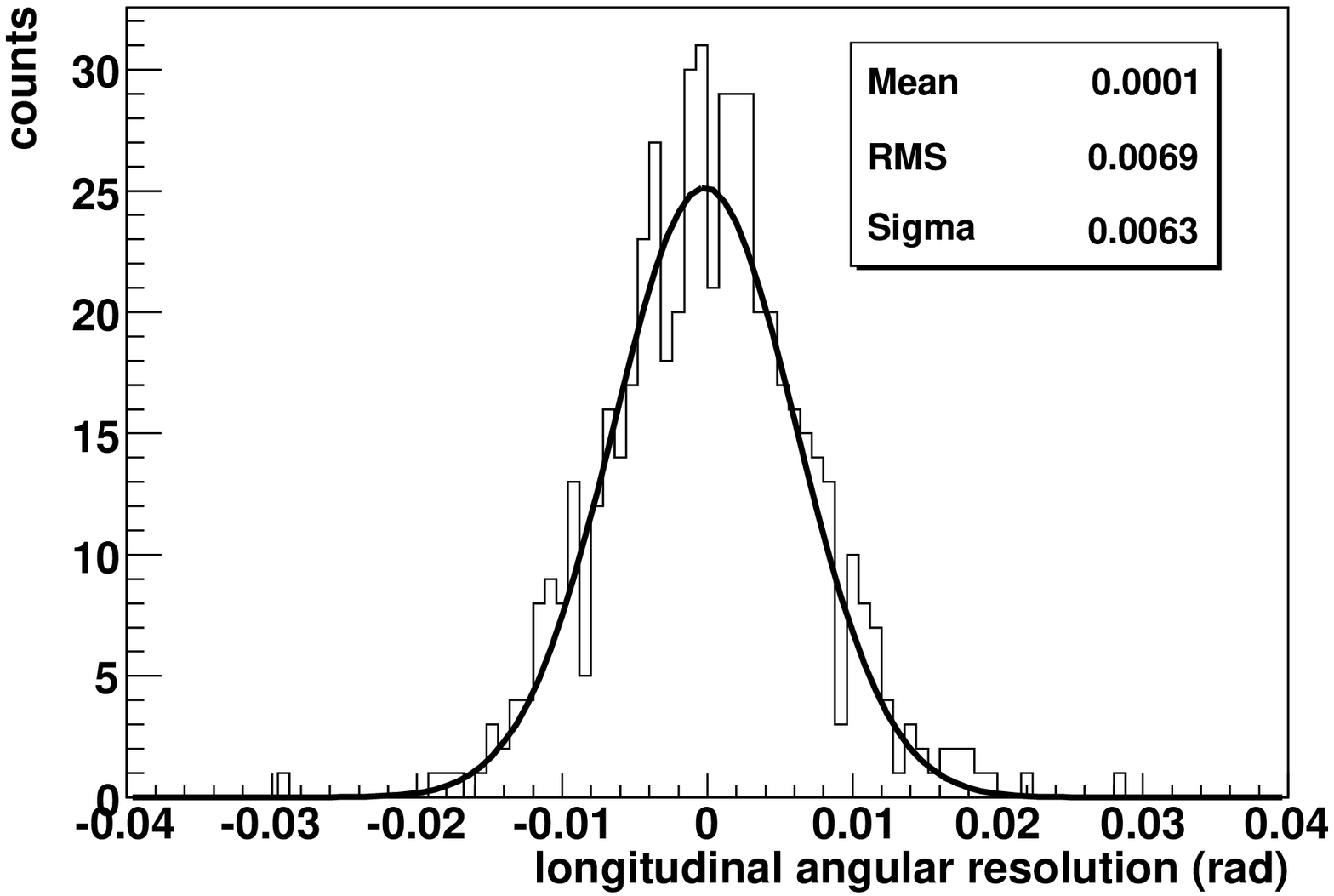,scale=0.35} \\
\end{tabular}

\caption{Angular residuals between base-tracks and corresponding volume tracks for particles crossing the emulsions
perpendicularly (left) and at $600 \, \rm mrad$ inclination (right). \label{fig:vt_sl}}
\end{center}
\end{figure*}

The angular residuals, computed as differences between base-track and volume-track angles, are shown in
Fig. \ref{fig:vt_sl}. Perpendicular tracks (left) can be reconstructed with a resolution of about
$1.7 \, \rm mrad$; the resolution worsens with the angle up to $6.3 \, \rm mrad$ for tracks crossing the brick
at about $600 \, \rm mrad$ with respect to the perpendicular direction (right).

\begin{figure}[]
\vspace{25pt}
\begin{center}
\epsfig{file=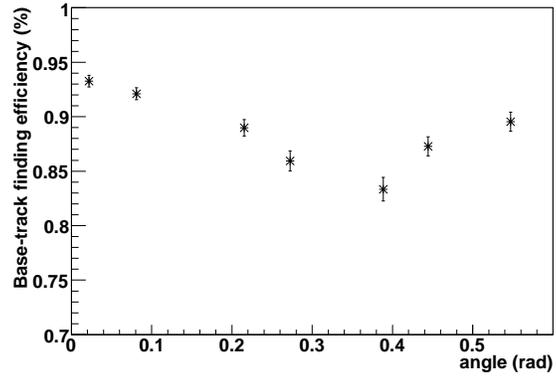,scale=0.4,bb=0 0 470 290}
\caption{Base-track finding efficiency as a function of track angle. \label{fig:eff}}
\vspace*{30pt}
\end{center}
\end{figure}

In order to evaluate the tracking efficiency, minimizing the effect of base-tracks not related to the beam,
sets of six adjacent films were aligned and beam volume-tracks\footnote{Beam tracks were defined as those
with a reconstructed angle within $3 \, \rm \sigma$ from the center of any of the 7 beam peaks.}
with at least five consecutive base-track segments were selected. Set by set, the efficiency was computed
for the two external films as the ratio between the number of tracks with six measured segments
and the number of tracks with five measured base-tracks only.
By averaging over several films, the plot in Fig. \ref{fig:eff} was obtained.
The base-track finding efficiency is larger than $90 \%$ for perpendicular tracks; correspondingly,
the micro-tracking efficiency, as given by its square root\footnote{Micro-tracks in the two emulsion layers
of each film are independently recognised and the efficiency of the linking procedure can be assumed equal
to 1 since it only depends on geometrical acceptances that can be properly tuned.}, is above $95 \%$.

\begin{figure}[]
\begin{center}
\epsfig{file=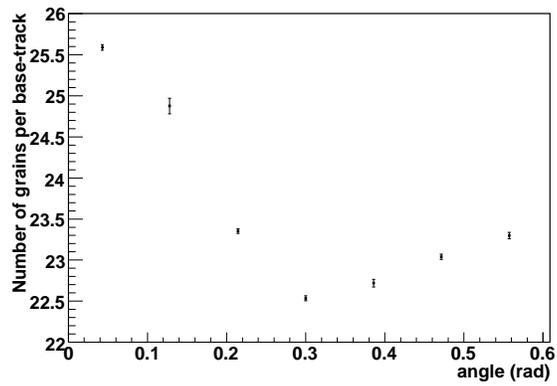,scale=0.4,bb=0 0 470 290}
\caption{Average number of grains per base-track as a function of track angle. \label{fig:grains}}
\end{center}
\end{figure}

The shape of the curve in Fig. \ref{fig:eff} is related to the slope dependence of the average number of grains
per base-track shown in Fig. \ref{fig:grains}: for small angles, the \emph{shadowing} effect (see \cite{ESS}) enhances
the signal, thus improving the efficiency.
The effect falls off with increasing slope. However, for sufficiently large angles,
the increase in the average number of grains due to longer path length in emulsion produces again an increase of the
efficiency.

The background measurement was performed using unexposed reference films refreshed and developed
at the same time as the other films.
A few hundred base-tracks selected applying standard cuts (see \cite{ESS})
were visually inspected in order to separate the contribution due to residual cosmic rays from instrumental background,
consisting of base-tracks with at least one micro-track generated by accidental combinations of fog grains.
About $1$ fake base-track $/ \rm cm^2$ was found in the angular range $[0,400] \, \rm mrad$.

The measurements presented in this section were performed with the ESS working at its maximum speed of 
$\sim \rm 20 \, cm^2 / h$.

\section{Particle track reconstruction in OPERA bricks}

Two different steps are foreseen in the event location strategy we plan to apply for the analysis of
the OPERA bricks.
The first procedure is the so-called \emph{scan-back}: starting from a set of predictions provided by the
electronic detectors, tracks of secondary particles coming from a neutrino interaction are searched for
in the CS films;
then, these tracks are followed back, film by film, from the most downstream emulsion of the brick
to the interaction point where they originate.
Whenever a track disappearance signal is detected (the track is not found in a certain number of consecutive 
films set according to the tracking efficiency), an area of about $25 \, \rm mm^2$ is measured in several films 
around the candidate vertex point (\emph{volume scan}) in order to fully reconstruct the event and 
apply topological and kinematical selection criteria.

In order to test the system capability in reconstructing particle tracks in the OPERA emulsion-lead target,
several bricks assembled with \emph{refreshed} films were exposed to a low intensity $8 \, \rm GeV/\it c$ $\pi^-$ beam
at the CERN PS - T7 line.
CS films, packed under vacuum in a separate envelope, were attached to the downstream face of each brick. \\
About $1000$ pions with inclination angles of about $50 \, \rm mrad$ with respect to the perpendicular
direction were recorded in each target unit over an area of $9 \times 9 \, \rm cm^2$.
Before unpacking and development, CS films were removed\footnote{Since the scanning of the CS films is a crucial
step in the event location procedure (only tracks measured in the CS films are followed-back across the brick),
the background track density should be kept as low as possible.} and each brick was exposed to cosmic rays
for $6$ hours in order to collect a set of passing-through tracks (density \mbox{$\sim 1$ track $/ \rm mm^2$})
to be used for film alignment. 

The OPERA event location procedure starting from CS films was mimicked using the tracks found in both CS's 
and confirmed in the most downstream film of the brick as predictions.
Film by film, each track was searched for in one single microscope view ($390 \times 310 \, \rm \mu m^2$)
within a slope-dependent angular window, accounting for the decrease in accuracy with increasing angle.
In case two or more base-tracks satisfied the angular cut, the closest-distance criterion was used
for selection\footnote{Different best-candidate selection criteria (for example, criteria based on the 
angular agreement) are currently being tested.}. 
Track propagation from film to film was performed by using the position and the slope of 
the most upstream track segment.
Losses due to tracking inefficiency were recovered by requiring that the search for each track be iterated
in 3 consecutive emulsion films if no candidate was found in one film. In case no base-track
compatible with the prediction was found in 3 films, the position of the most upstream measured base-track
was tagged as a candidate vertex point and the volume scan procedure was applied around it.

\begin{figure}[t!]
\begin{center}
\epsfig{file=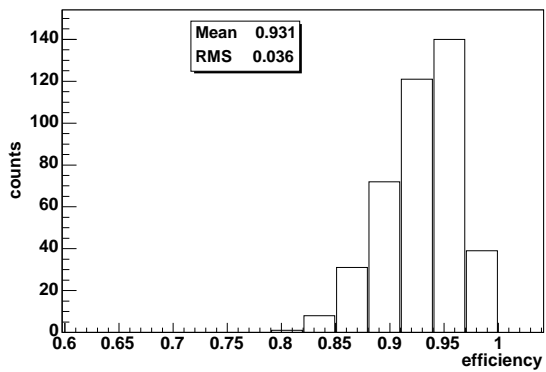,scale=0.4}
\caption{Tracking efficiency for non-interacting particles. \label{fig:sb_eff_passing}}
\end{center}
\end{figure}

\begin{figure*}[ht!]
\begin{center}
\begin{tabular}{cc}
    \epsfig{file=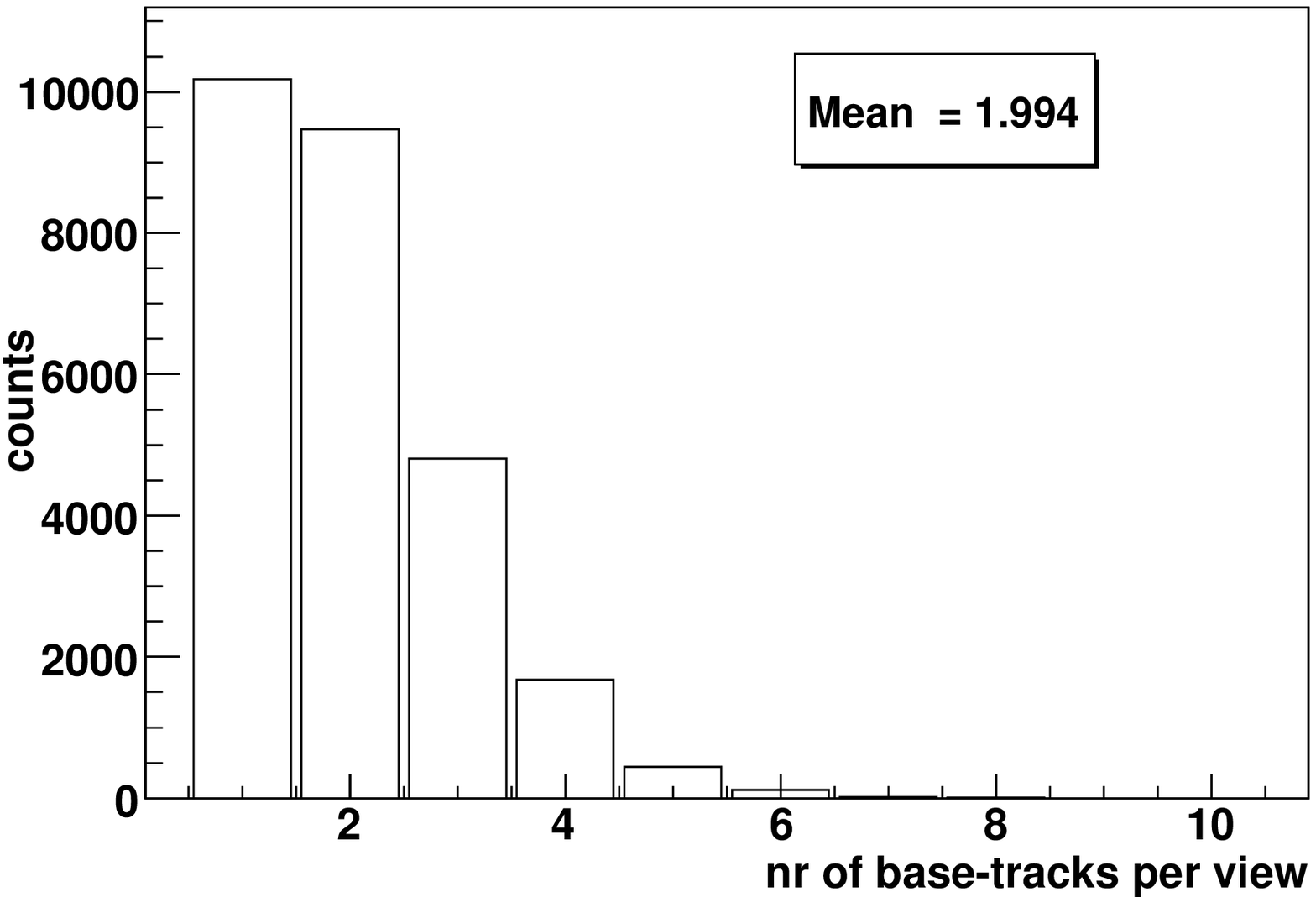,scale=0.35}
        & \epsfig{file=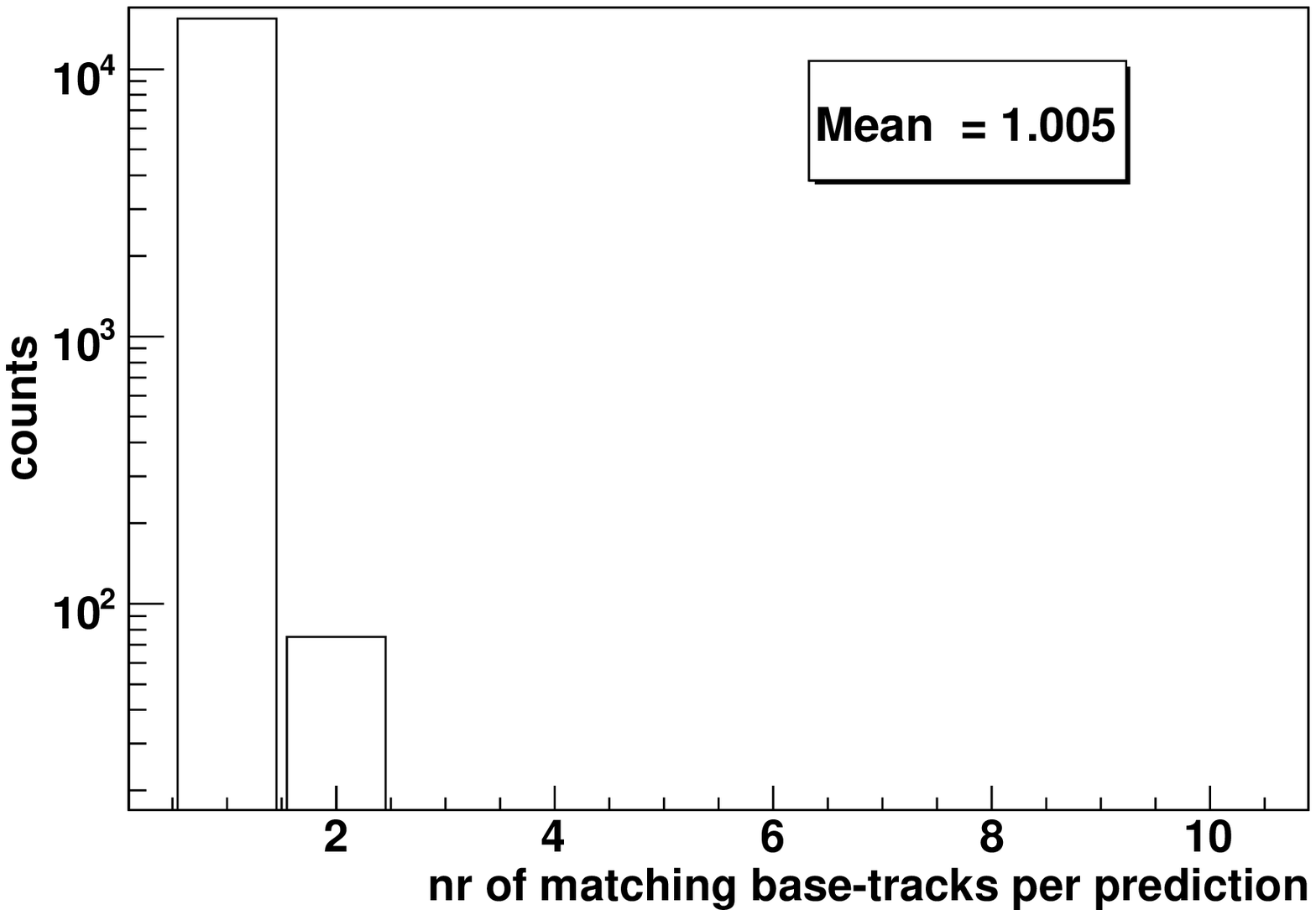,scale=0.35} \\
\end{tabular}

\caption{Left: Number of base-tracks per view. Right: Number of base-tracks per prediction satisfying the selection criteria.
    \label{fig:base_pred}}
\end{center}
\end{figure*}

\begin{figure}[hc]
\begin{center}
\epsfig{file=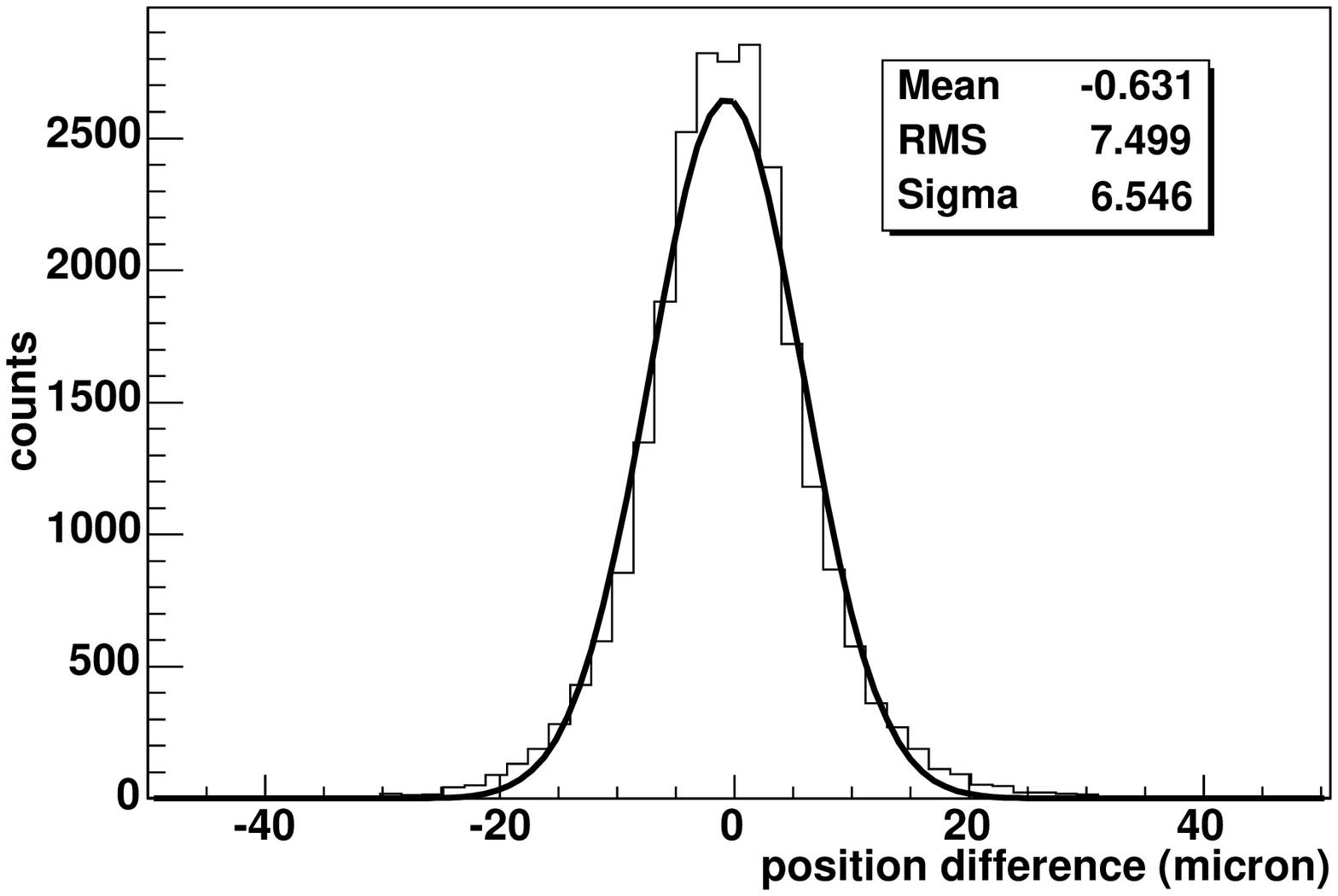,scale=0.4}
\caption{Position differences between predicted and found base-track segments. 
    Tracks  are distributed over an area of about $40 \, \rm cm^2$.
    \label{fig:sb_posdiff_found_pred}}
\end{center}
\end{figure}

The tracking efficiency, computed as the number of found base-tracks divided by the total number of
films for the sample of passing-through particles (mainly primary pions and muons from beam
contamination), is shown in Fig. \ref{fig:sb_eff_passing}. The measured value ($\sim 93 \%$) is in
agreement with the expected behaviour at small angle shown in Fig. \ref{fig:eff}.

On average, 2 base-tracks with space angles smaller than $400 \, \rm mrad$ were reconstructed
in each microscope view around the position of a given predicted track (Fig. \ref{fig:base_pred}, left).
By requiring that the slope difference between found and predicted base-tracks be not larger than 
$( 0.03 + 0.05 \times \rm slope )$, hence allowing for the non-negligible contribution due to the scattering 
of low energy particles in lead, the average number of candidates per prediction
reduces to 1, the fraction of predictions with multiple candidates being of the order of $10^{-3}$
(Fig. \ref{fig:base_pred}, right).

\begin{figure*}[ht!]
\begin{center}
\begin{tabular}{cc}
    \epsfig{file=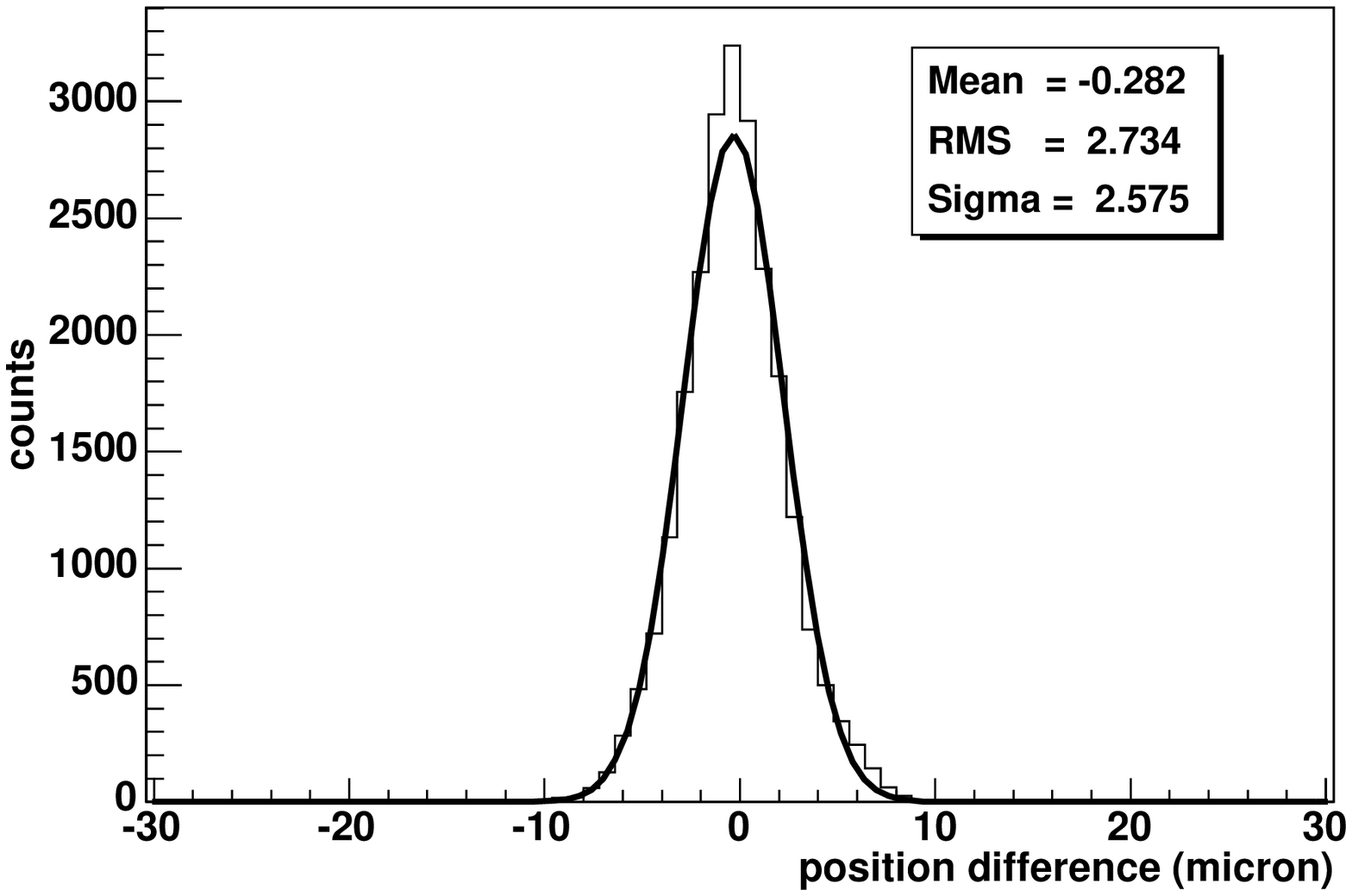,scale=0.35}
        & \epsfig{file=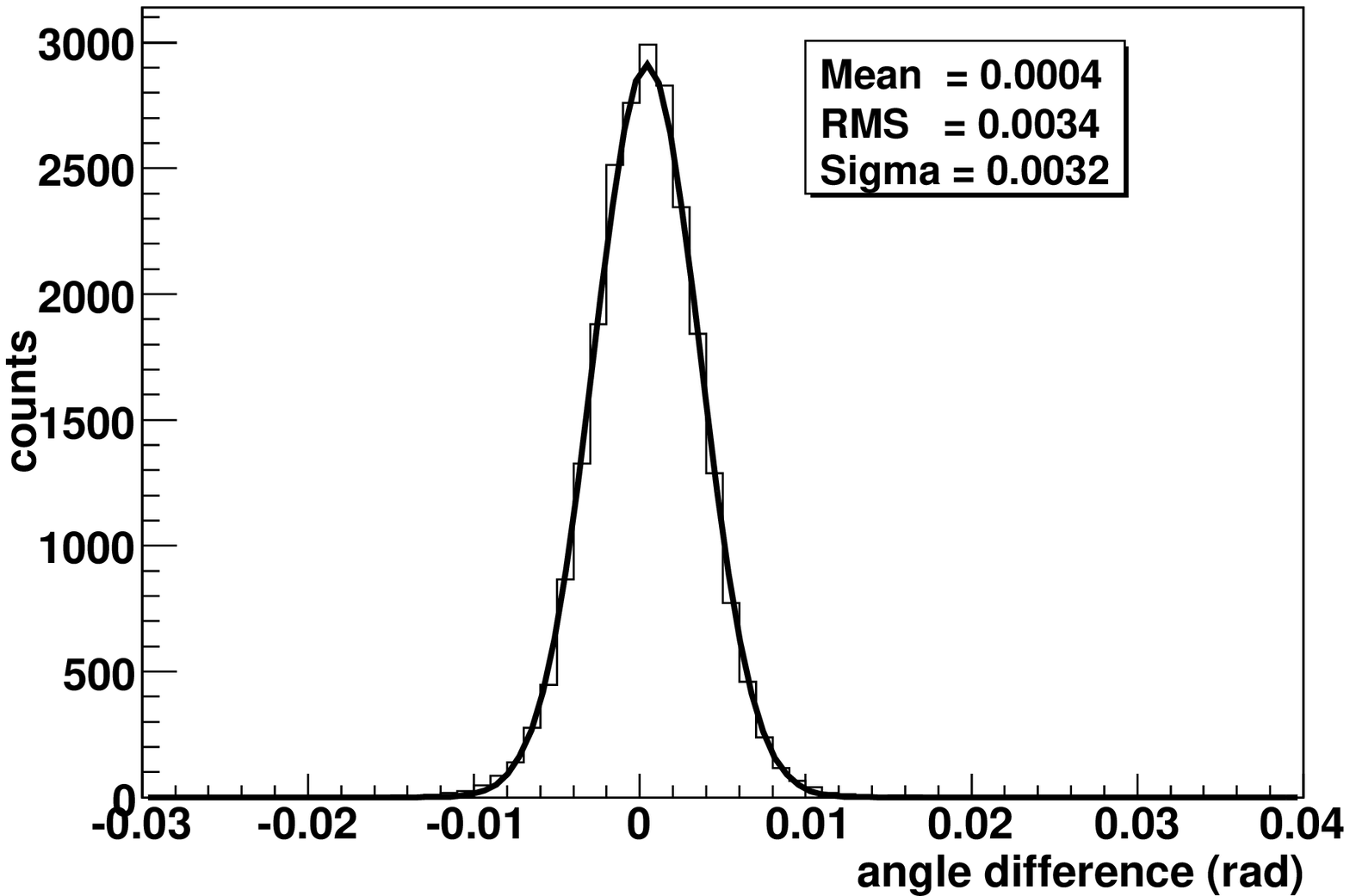,scale=0.35} \\
\end{tabular}

\caption{Position (left) and angle (right) differences between base-tracks selected in two scan-back
procedures corresponding to the same set of predictions. \label{fig:reproduce}}
\end{center}
\end{figure*}

We foresee to complete the scanning of each brick in a few hours. 
In order to minimise the scanning time required for event location, fine corrections are not applied 
in scan-back mode.
First of all, only a \emph{global} film inter-calibration procedure is performed. 
Fig. \ref{fig:sb_posdiff_found_pred} shows the distribution of the position differences between predicted and 
found base-track segments. The plot refers to tracks scattered over an area of about $40 \, \rm cm^2$. 
The sigma of the gaussian fit is of about $7 \, \rm \mu m$.
Moreover, only one microscope view is measured for each predicted track, thus effects such as local
planarity cannot be corrected. The resulting angular precision is of about $4 \div 5 \, \rm mrad$.

The scan-back reproducibility was studied by performing the procedure twice for the same set of predictions
and by comparing film by film the selected base-tracks per each predicted track. As shown in Fig. \ref{fig:reproduce},
the reproducibility is of about $2.6 \, \rm \mu m$ in position and $3 \, \rm mrad$ in angle.

As a cross-check for the validation of scan-back results,
the volume scan procedure was applied around track disappearance points
and, by a combined analysis of scan-back and volume scan data, events were  classified as
single and multi-prong interactions, low energy (scattering) tracks and passing-through tracks. 
Film-dependent effects, such as local distortion and misalignments resulting from large rotations, 
not completely corrected in the inter-calibration procedure, were found to be the main sources of 
inefficiency ($< 5 \%$) in the track following. Such effects can be accounted for by enlarging the 
search window in case no base-tracks compatible with a given prediction are found and by implementing 
a \emph{prediction-forking} method, allowing for multiple candidate following, if the slope/position 
acceptances are increased. 
Failures of the system in passing-through track following can be thus reduced to $\sim 1 \%$.

\section{Conclusions}
The particle tracking performance of the European Scanning System (ESS), a novel high-speed 
automatic microscope for the measurement of nuclear emulsion films, has been presented. 

The system, developed for the location and analysis of neutrino interactions in the emulsion-lead target 
of the OPERA experiment, fully satisfies the requirements of high tracking efficiency and accuracy, 
working at a speed of $\sim \rm 20 \, cm^2 / h$. 

The results of a test exposure to high-energy pions, designed to study the methods to be applied for film-by-film alignment, 
track following and event location and confirmation, were discussed in details, showing that the system is able to track 
particles produced in beam interactions with lead through the whole brick and identify the vertex points where they originate. 
The achieved results make us confident in view of forthcoming CNGS neutrino runs.

\section*{Acknowledgements}
We acknowledge the cooperation of the members of the OPERA Collaboration and we thank many colleagues
for discussions and suggestions.
We gratefully acknowledge the invaluable support of the technical staff in our laboratories;
in particular, we thank A. Andriani, P. Calligola, P. Di Pinto, M. Hess, H. U. Schuetz, C. Valieri
for their contributions.
We acknowledge support from our funding agencies.
We thank INFN for providing fellowships and grants (FAI) for non Italian citizens.



\end{document}